\documentclass[reprint, aps, prl, showpacs, floatfix, 
longbibliography]{revtex4-1}

\usepackage[T1]{fontenc}
\usepackage{amsmath}
\usepackage{accents}
\usepackage{textcomp}
\usepackage{tgtermes}
\usepackage[altg]{qtxmath}
\renewcommand*{\vec}[1]{\boldsymbol{#1}}
\usepackage{microtype}
\usepackage{graphicx}
\usepackage{braket}
\usepackage{blindtext}
\usepackage[usenames,dvipsnames]{color}
\graphicspath{{./}}
\usepackage[breaklinks=true,
pdftitle={}]{hyperref}
\usepackage{filemod}
\usepackage{datetime}
\usdate

\usepackage{mathrsfs}
\usepackage{amsbsy}

\newcommand*{\D}{\mathrm{d}}
\newcommand*{\e}{\mathrm{e}}

\makeatletter
\AtBeginDocument{\renewcommand{\@biblabel}[1]{#1.}}
\def\@bibdataout@aps{%
 \immediate\write\@bibdataout{%
  @CONTROL{%
   apsrev41Control%
   \longbibliography@sw{%
    ,author="48",editor="1",pages="1",title="0",year="0"%
   }{%
    ,author="0a",editor="1",pages="0",title="",year="1"%
   }%
  }%
 }%
 \if@filesw
  \immediate\write\@auxout{\string\citation{apsrev41Control}}%
 \fi
}%
\makeatother

\begin{document}

\title{Polarized laser-wakefield-accelerated kiloampere electron beams}

\author{Meng Wen}
\email{meng.wen@mpi-hd.mpg.de}
\affiliation{Max-Planck-Institut f{\"u}r Kernphysik, Saupfercheckweg~1, 69117~Heidelberg, Germany}
\author{Matteo Tamburini}
\email{matteo.tamburini@mpi-hd.mpg.de}
\affiliation{Max-Planck-Institut f{\"u}r Kernphysik, Saupfercheckweg~1, 69117~Heidelberg, Germany}
\author{Christoph H. Keitel}
\affiliation{Max-Planck-Institut f{\"u}r Kernphysik, Saupfercheckweg~1, 69117~Heidelberg, Germany}

\date{\today}

\begin{abstract}
High-flux polarized particle beams are of critical importance for the investigation of spin-dependent processes, such as in searches of physics beyond the Standard Model, as well as for scrutinizing the structure of solids and surfaces in material science. Here we demonstrate that kiloampere polarized electron beams can be produced via laser-wakefield acceleration from a gas target. A simple theoretical model for determining the electron beam polarization is presented and supported with self-consistent three-dimensional particle-in-cell simulations that incorporate the spin dynamics. By appropriately choosing the laser and gas parameters, we show that the depolarization of electrons  induced by the laser-wakefield-acceleration process can be as low as~10\%. Compared to currently available sources of polarized electron beams, the flux is increased by four orders of magnitude.
\end{abstract}

\maketitle

Polarized beams of electrons, photons and positrons are widely employed in materials science~\cite{Schultz:1988:RevModPhys.60.701} as well as in atomic, nuclear and particle physics~\cite{Tolhoek:1956:RevModPhys.28.277, Danielson:2015:RevModPhys.87.247} because they allow the investigation both of the spin dependence of fundamental interactions and of the violation of symmetries such as parity~\cite{Schlimme:2013:PhysRevLett.111.132504}. In fact, precise measurements of spin-dependent processes enable searches of physics beyond the Standard Model that can compete with direct searches at high-energy accelerators~\cite{QweakN2018}. In addition to their intrinsic interest, polarized electron beams can also be employed to generate polarized beams of photons~\cite{PhysRevLett.108.264801} and positrons~\cite{PhysRevLett.116.214801} which, in turn, can be employed to address long-standing problems such as the matter-antimatter asymmetry in the Universe.

Currently, polarized electron beams are mainly produced in storage rings via radiative polarization due to the Sokolov-Ternov effect~\cite{Sokolov:SPJ:1967} or by extracting polarized electrons~\cite{Kessler:1985:Polarized-Electrons} directly from polarized atoms and polarized photocathodes~\cite{Pierce:apl:1975}. However, the maximal electric current of polarized electron beams both from storage rings~\cite{Barber:NIMPRSA:1993, Sun:2010:NIMPRSA339} and from photocathodes~\cite{Drachenfels:aipcp:2003, Maxson:RSI:2014, Hernandez-Garcia:IEEE:2016} is limited to less than $10^{-1}$~ampere due to their operational voltages. Other methods based on spin filters~\cite{Batelaan:1999:PhysRevLett.82.4216}, Stern-Gerlach-like beam splitters~\cite{Dellweg:2017:PhysRevLett.118.070403}, and radiative polarization~\cite{Li:PhysRevLett:2019} also yield relatively low currents. The low attainable electric currents noticeably limit experiments with polarized beams~\cite{Schultz:1988:RevModPhys.60.701, Tolhoek:1956:RevModPhys.28.277, Danielson:2015:RevModPhys.87.247} as well as the flux and the brightness of the polarized photon and positron beams attainable from polarized electron beams~\cite{PhysRevLett.108.264801, PhysRevLett.116.214801}.

\begin{figure}[t]
\centering
\includegraphics[width=\linewidth]{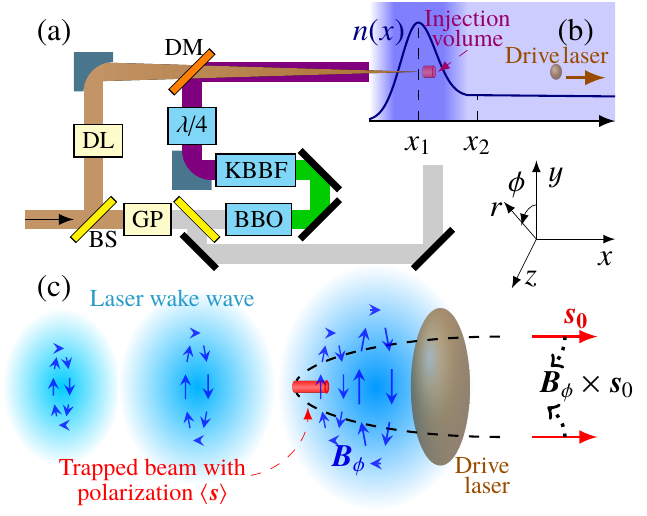}
\caption{(a)~Schematic layout. The femtosecond/picosencond optical laser (brown/gray stripe) and the UV laser (purple stripe). BS: beam splitter, DL: delay line, DM: dichroic mirror, GP: grating pair, BBO and KBBF crystals, $\lambda/4$: quarter-wave plate. (b)~The longitudinal plasma density profile $n(x)$. A density bump centered at $x_1$ is followed by a uniform density plasma for $x\geq x_2$. The purple cylinder near the plasma density peak indicates the electrons of the injection volume, the brown ellipsoid  the optical laser pulse driving the wakefield. (c)~Illustration of the wake wave with its azimuthal magnetic field $\vec{B}_\phi$. Dashed black lines represent two possible trajectories of off-axis electrons, the dotted black arrows show the precession of their spin, $s_0$ is the initial spin.}
\label{fig:scheme}
\end{figure}

Here we put forward a method for generating polarized electron beams with currents \emph{four} orders of magnitude larger than those attainable with currently existing methods. It consists in the rapid electron polarization of a gas jet via photodissociation by a circularly-polarized UV laser pulse followed by electron laser-wakefield acceleration by an optical laser pulse. An illustration of our scheme is displayed in Fig.~\ref{fig:scheme}. A tens femtoseconds linearly polarized optical laser pulse with hundreds mJ energy and low intensity is divided into two pulses with a beam splitter. One of the two pulses first passes through a grating pair, where it is stretched to hundreds picoseconds, and is further divided into two pulses with a second beam splitter. One picosecond pulse is used to initially align the molecular bonds before dissociation~\cite{Friedrich1995, Sofikitis:2017:PhysRevLett.118.233401, huetzenHPLSE19}, the other one undergoes frequency quadruplication with a BBO and KBBF crystal, and its bandwidth is tightened via reflection with several narrow band coated mirrors [not displayed in Fig.~\ref{fig:scheme}(a)]. The generated UV laser pulse is converted to circular polarization with a quarter-wave plate and focused to the gas target where it polarizes electrons along the laser propagation direction via molecular photodissociation~\cite{Rakitzis:Science:2003:1936, Rakitzis:CPC:2004, Sofikitis:EPL:2008, Sofikitis:2017:PhysRevLett.118.233401, Sofikitis:2018:PhysRevLett.121.083001, huetzenHPLSE19}.
Note that electronically polarized atomic hydrogen with density up to $10^{19}\text{ cm$^{-3}$}$ was already experimentally obtained by employing laser-induced molecular photodissociation~\cite{Sofikitis:2018:PhysRevLett.121.083001}. The second of the two pulses from the first beam splitter passes through a controllable delay line and reaches the previously generated electronically polarized gas target to drive wakefield acceleration [see Fig.~\ref{fig:scheme}(a)]. The delay between the polarizing UV laser pulse and the optical laser pulse driving wakefield acceleration must be much smaller than $\sim 1$~ns, i.e., the time needed to the hyperfine coupling to transfer polarization from electrons to nuclei~\cite{Sofikitis:2017:PhysRevLett.118.233401}. 
According to the shock-front injection method~\cite{PhysRevLett.110.185006}, the electron density $n(x)$ is tailored along the laser propagation direction $x$ [see Fig.~\ref{fig:scheme}(b)]. This allows to control the local plasma period (wavelength) $\tau_p(x) = 2 \pi \sqrt{\varepsilon_0 m/e^2 n(x)}$ [$\lambda_p(x) = c \tau_p(x)$] along the laser propagation direction such that only the electrons in the downramp of the density, i.e. in the region between $x_1$ and $x_2$ of Fig.~\ref{fig:scheme}(b), are injected and accelerated into the wake wave~\cite{PhysRevLett.110.185006}. Here $\varepsilon_0$ is the vacuum permittivity, $m$ and $-e$ are the electron mass and charge, respectively.

The initial electronic polarization (IEP) of the gas target $P_0$ depends on the employed molecular species. In particular, hydrogen halides were used to generate dense electronically polarized hydrogen and deuterium atoms~\cite{Sofikitis:2017:PhysRevLett.118.233401, Sofikitis:2018:PhysRevLett.121.083001}. In this case, the presence of other species than hydrogen implies that, although the IEP of hydrogen can reach 100\% \cite{Sofikitis:2017:PhysRevLett.118.233401}, the global IEP of the generated plasma is less than unity. However, as it will be clear in the following, only  optical laser pulses with intensity of the order of $10^{18}\text{ W/cm$^2$}$ drive laser-wakefield acceleration without significant electron beam depolarization. Hence, for halogen atoms only the outer shell electrons have appreciable probability to be extracted and accelerated~\cite{Clayton:2010:PhysRevLett.105.105003,Chen:JCP:2013}. By considering HF molecules, six paired electrons in the outer-shell of fluorine may be extracted and accelerated by the optical driver laser pulse, which results into an IEP $P_0\approx25\%$. 
Higher IEP is expected by employing molecular hydrogen. However, single photon dissociation of molecular hydrogen occurs at wavelength $<100\text{ nm}~$\cite{Glass-Maujean:1988:PhysRevLett.61.157}, which has not yet been realized experimentally but may be attained via laser-plasma techniques~\cite{Chen:NC:2016}.

In the following, we investigate electron injection and acceleration in the case of an initially fully electronically polarized plasma with spin directed along the driver laser propagation direction~$x$. In fact, a partially polarized gas is a mixture of a fully polarized fraction $P_\mathrm{0}$ and of an unpolarized fraction $(1-P_\mathrm{0})$~\cite{Kessler:1985:Polarized-Electrons}. Since the unpolarized fraction remains unpolarized, the final electron beam polarization (EBP) from a partially polarized gas is $P_\mathrm{fin}=P_\mathrm{0} P$, where $P$ is the final EBP of an initially fully polarized plasma. Depolarization of an initially polarized target occurs due to the electromagnetic field inhomogeneities in the wake of the driving laser pulse.  
In particular, the spin of injected off-axis electrons precesses differently depending on the electron position, and mainly according to the structure of the azimuthal magnetic field in the wake wave~\cite{Gorbunov:1996:PhysRevLett.76.2495, Gorbunov:POP:1997, Andreev:PPR:1997} [see Fig.~\ref{fig:scheme}(c)]. As it will be clear below, the IEP is preserved in the weakly-nonlinear wakefield acceleration regime~\cite{Gorbunov:1987,Esarey:1997:PhysRevLett.79.2682,Andreev:POP:1997,Marques:POP:1998,Gorbunov:Physics_of_Plasmas:2005,Kalmykov:2006:POP}, with most of the injected electrons moving predominantly along the laser propagation direction and originating from the region near the laser propagation axis.

The electron beam dynamics is investigated with fully three-dimensional particle-in-cell (PIC) simulations with the code EPOCH~\cite{Arber:epoch:2015hc}. In addition to the electron position $(x,y,z)$ and velocity $(v_x, v_y, v_z)$, we implemented in the code the electron spin dynamics~\cite{Wen:2017:PhysRevA.95.042102}. Following Ehrenfest's theorem~\cite{Ehrenfest1927}, a vector $\vec{s}=\braket{\Phi |\vec{\sigma}|\Phi}$ with $\left| \vec{s} \right|=1$ is used to describe the spin of an electron in a quasiclassical state, where $\Phi$ denotes the normalized two-component spinor and $\vec{\sigma}$ are the Pauli matrices \cite{Mane:RPP:2005}.  In general, the EBP is defined as the statistical average over all electrons of the beam \mbox{$P=\left|\sum_{i}^N \vec{s}_i /N\right|$}, where $N$ is the total number of electrons of the beam.
The spin of an electron in an electric $\vec{E}$ and magnetic $\vec{B}$ field precesses according to the Thomas-Bargmann-Michel-Telegdi equation $\D\vec{s}/\D t = (\vec{\Omega}_\mathrm{T}+\vec{\Omega}_\mathrm{a})\times \vec{s}$ with

\begin{subequations}
\label{eq:dsdt}
  \begin{align}
  \label{eq:dsdtT}
    \vec{\Omega}_\mathrm{T} & = \frac{e}{m}\left(\frac{1}{\gamma}\vec{B} -\frac{1}{\gamma+1}\frac{\vec{v}}{c^2}\times\vec{E}\right) \,, \\
    \vec{\Omega}_\mathrm{a} & = a_\e \frac{e}{m}\left[\vec{B} - \frac{\gamma}{\gamma+1}\frac{\vec{v}}{c^2}\left(\vec{v}\cdot\vec{B}\right) -\frac{\vec{v}}{c^2}\times\vec{E}\right]\,,
  \end{align}
\end{subequations}
where $a_\e \approx 1.16\times10^{-3}$ is the anomalous magnetic moment of the electron. Note that electrostatic Coulomb collisions do not affect the electron spin, such that depolarization via collisions can only occur due to spin-orbit and spin-spin interaction~\cite{Berestetskii-book,Kulsrud:1982:PhysRevLett.49.1248}. Indeed, for $10^{19}\text{ cm$^{-3}$}$ spin polarized hydrogen density obtained from hydrogen halide molecular dissociation, a collisional polarization lifetime larger than 1~ns was demonstrated~\cite{Sofikitis:2018:PhysRevLett.121.083001}, which is much longer than the electron injection and acceleration time of approximately 1.1~ps considered here (see below). Also, Stern-Gerlach and radiation reaction forces are negligible at the low intensities and copropagating geometry considered here~\cite{Wen:2017:PhysRevA.95.042102,tamburiniNJP10}.

In our PIC simulations we have chosen laser and plasma parameters similar to those already obtained in experiments with shock-front injection~\cite{PhysRevSTAB.13.091301, PhysRevLett.110.185006, swanson17, Xu:POP:2017}. The plasma density profile is shown in Fig.~\ref{fig:scheme}(b). A density bump centered at $x_1=50\lambda$ with peak density $n_1=4\times10^{-3} n_\mathrm{c}$ is followed by a plateau beginning at $x_2=60\lambda$ and having uniform density $n_2= 10^{-3} n_\mathrm{c}$, where $n_\mathrm{c} = \varepsilon_0 m \omega^2 / e^2$ is the critical plasma density, $\lambda=0.8\text{ $\mu$m}$, $\tau=\lambda/c$, and $\omega=2\pi/ \tau$ are the laser wavelength, period, and frequency, respectively. At $t=0$ a laser pulse linearly polarized along the $y$-axis and propagating along the positive $x$-axis enters the simulation box from $x=0$ with focus located at $x_1$. The laser pulse has Gaussian envelope $a_0 \exp[-(t - x/c - 2 T_l)^2/T_l^2 - r^2/w_0^2]$, where $w_0=10\lambda$, $T_l=8\tau$, $a_0 = e E_0 / m \omega c$ is the normalized laser field amplitude, and $E_0$ is the maximal laser field. A moving window was employed, with $150\lambda(x)\times 60\lambda(y)\times 60\lambda(z)$ computational box and $3000(x)\times 450(y)\times 450(z)$ grid points.

\begin{figure}[t]
\centering
\includegraphics[scale=0.7]{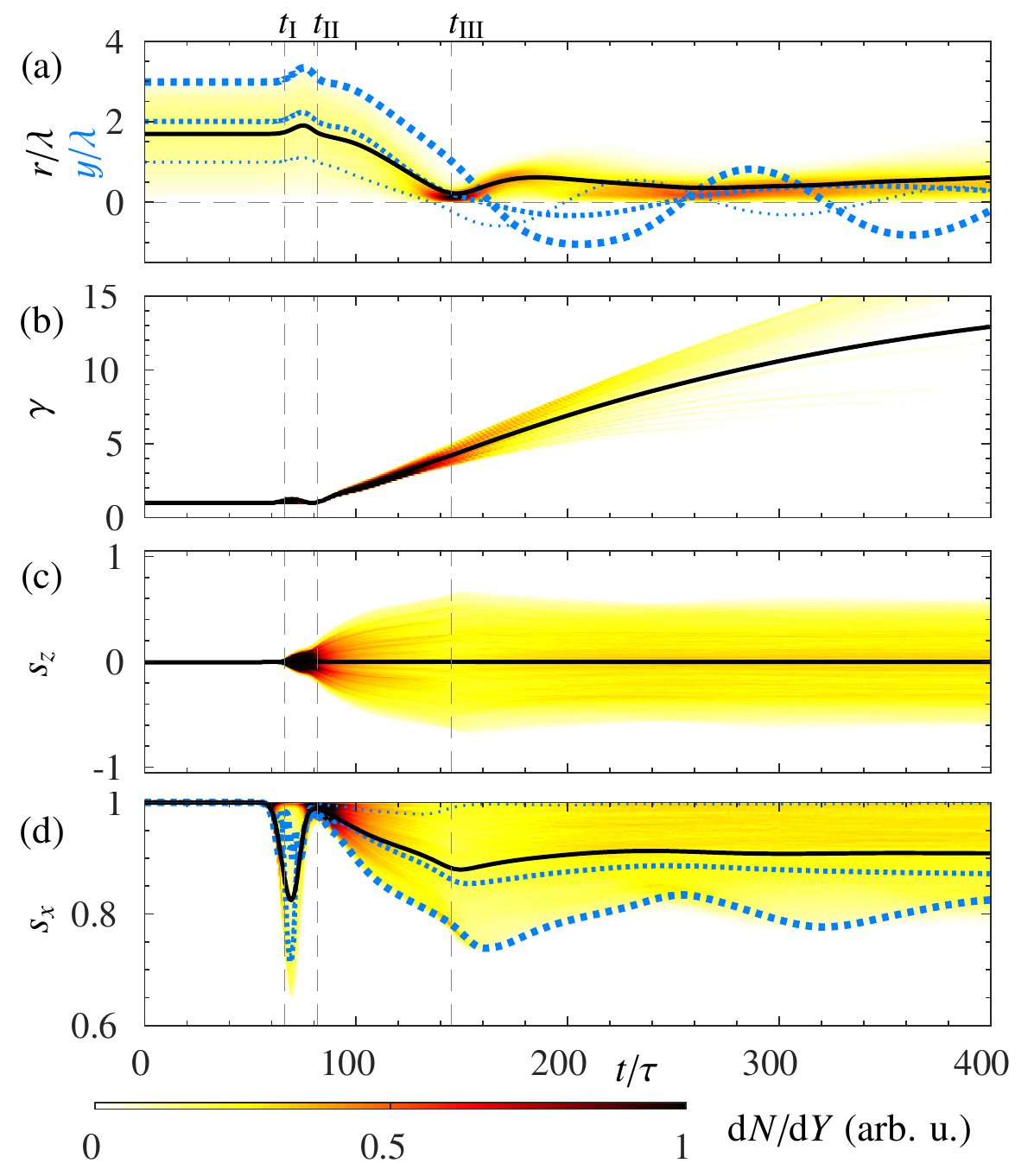}
\caption{Injected electron distributions as a function of time in a weakly nonlinear wake driven by an optical laser pulse with $a_0=1$. (a)~Radial coordinate, (b)~relativistic factor, (c)~the $z$ component of the spin, (d)~the $x$ component of the spin. The dotted light-blue lines represent the evolution of three electrons with initial coordinate $y_0=(1,2,3)\lambda$. The solid black line shows the evolution of the statistical average of the reported quantity  $\braket{Y}$, where $Y = (r/\lambda,\gamma,s_z,s_x)$ in panels~(a)-(d), respectively. The colorbar displays the scale of $\D N/\D Y$.} \label{fig:x250}
\end{figure}

Figure~\ref{fig:x250} reports the evolution of the distribution of
trapped electrons for a laser pulse with $a_0=1$ as a function of the radial coordinate $r$ [Fig.~\ref{fig:x250}(a)], of the relativistic factor $\gamma=1/\sqrt{1-v^2/c^2}$ [Fig.~\ref{fig:x250}(b)], of the $z$ [Fig.~\ref{fig:x250}(c)] and $x$ [Fig.~\ref{fig:x250}(d)] components of the electron spin $\vec{s}$. In addition, the dotted light-blue lines in Fig.~\ref{fig:x250} display the evolution of three electrons with initial coordinate $y_0=(1,2,3)\lambda$.
The evolution is divided into four stages. Stage~I corresponds to $t \leq t_\mathrm{I}=x_1/c+2T_l=66\tau$, i.e. the time before the laser pulse peak reaches the injection volume, which is located near the maximum of the density bump $x_1$. In fact, after the local plasma wavelength $\lambda_p(x)$ reaches its minimum $\lambda_p(x_1)$, the density downramp leads to the expansion of the wake with a decrease of the wake phase velocity, which results into electron injection in the wakefield~\cite{PhysRevLett.110.185006}. The injection region ends before the uniform density plasma at $x_2$, i.e. when the local plasma wavelength reaches $\lambda_p(x_2)$. In our simulations, we trace all and only the electrons that remain in the accelerating phase of the wake. The initial region of trapped electrons constitutes the injection volume with a radius $R_\mathrm{inj} \propto w_0$~\cite{swanson17}. 

Stage~II begins at $t_\mathrm{I}$, where the electrons of the injection volume undergo a transient acceleration phase due both to the laser and to the wake-wave fields, as shown in Figs.~\ref{fig:x250}(a)-(b). This transient phase lasts approximately one plasma period corresponding to the local density at $x_1$, i.e. $t_\mathrm{II} = t_\mathrm{I} + \tau_p(x_1) \approx 82\tau$. During this stage the electrons of the injection volume move along the density downramp with mildly relativistic velocity. Note that at $t_\mathrm{II}$ the laser pulse peak is at $c(t_\mathrm{II} - 2T_l)\approx 66\lambda$, i.e. well inside the uniform plasma that starts at $x_2=60\lambda$, and that the oscillatory dynamics driven by the laser and the wake field tends to restore the initial momentum and spin of electrons (see Figs.~\ref{fig:x250}(a)-(d) at $t_\mathrm{II}$). Also, note that the linearly polarized laser pulse has its magnetic field directed along $z$, such that it does not directly affect $s_z$. Hence, the spreading of the $s_z$ distribution for $t>t_\mathrm{I}$ is determined by the azimuthal magnetic field of the wake wave [see  Figs.~\ref{fig:x250}(c)]. 

Stage~III starts at $t_\mathrm{II}$, where electron trapping with acceleration and focusing inside the wakefield occurs, as shown in Figs.~\ref{fig:x250}(a)-(b). The simultaneous processes of acceleration and focusing become manifest in the decrease of the radius $\braket{r}$ of the volume of the injected electrons [see Fig.~\ref{fig:x250}(a)] and in the increase of the electron beam relativistic factor $\braket{\gamma}$ [see Fig.~\ref{fig:x250}(b)], respectively. Here brackets denote the statistical average over the electrons of the injection volume. 
Simulations indicate that the focusing time, i.e. the time needed by the electrons of the injection volume to focus to the wake axis, is approximately $2\tau_p(x_2)$, which implies that stage~III ends at $t_\mathrm{III} = t_\mathrm{II} + 2\tau_p(x_2) \approx 145\tau$ (see Fig.~\ref{fig:x250}). Note that electron spin spreading occurs predominantly during stage~III, i.e. between $t_\mathrm{II}$ and $t_\mathrm{III}$, such that one can estimate the electron beam depolarization from the duration of stage~III and from the spin precession frequency during this stage (see below). 

During stage~IV ($t>t_\mathrm{III}$), the energy of the electron beam increases steadily, whereas the spin distribution and beam polarization remain stable [see Fig.~\ref{fig:x250}(c)-(d)]. 
Three factors determine the stability of the EBP during stage~IV. First, the constantly increasing relativistic factor $\gamma$ significantly reduces the precession frequency [see Eq.~\eqref{eq:dsdtT}]. Second, our simulations show that for $t>t_\mathrm{III}$ trapped electrons are predominantly focused and confined around the wakefield axis within a region of radius smaller than $\lambda$ [see Fig.~\ref{fig:x250}(a)]. This results into a suppressed precession frequency since transverse fields around the axis are relatively weak [see Eq.~(\ref{eq:Fr_r}) below]. 
Third, trapped electrons oscillate around the axis of the wakefield while accelerated [see the dotted light-blue lines in Fig.~\ref{fig:x250}(a)], which results into a change of the direction of $\vec{\Omega}_\mathrm{T} \propto \vec{B}_\phi$ at each crossing of the axis. Thus, $\D \vec{s} / \D t$ oscillates around zero, while $\braket{\D \vec{s} / \D t}$ averages out to zero. 

In order to evaluate the electron beam depolarization, we estimate the electron precession frequency during stage~III. Recalling that $|\vec{\Omega}_\mathrm{a}| \ll |\vec{\Omega}_\mathrm{T}|$ and taking the limit of slowly varying frequency $|\D\vec{\Omega}_\mathrm{T}/\D t| \ll |\vec{\Omega}_\mathrm{T}|^2$, at the end of stage~III the $x$ component of the spin of an electron initially directed along the $x$ axis is $s_x \approx \cos \left[\int^{t_\mathrm{III}}_{t_\mathrm{II}}{\D t |\vec{\Omega}_\mathrm{T}(t)|}\right]$. The precession frequency $|\vec{\Omega}_\mathrm{T}(t)|$ decreases steadily from its maximum at $t_\mathrm{II}$, where the injected electrons have $\gamma \approx 1$ and a relatively large $r$, which implies a larger $\vec{B}_\phi(r)$ (see Eq.~(\ref{eq:Fr_r}) below), to nearly zero at $t_\mathrm{III}$ (see the black curve of Fig.~\ref{fig:x250}(d) for $t \geq t_\mathrm{III}$). For simplicity, in our analytical estimate we assume a linear decrease of $|\vec{\Omega}_\mathrm{T}(t)|$ from $\vec{\Omega}_\mathrm{T}(t_\mathrm{II},r) \approx e\vec{B}_\phi(r)/(2m)$ at $t_\mathrm{II}$ to zero at $t_\mathrm{III}$ for all the electrons of the injection volume, which implies $s_x(r)\approx \cos \left[|\vec{\Omega}_\mathrm{T}(t_\mathrm{II},r)|(t_\mathrm{III} - t_\mathrm{II})/2\right]$. For $a_0\lesssim 1$, the quasistatic azimuthal magnetic field in the wake is~\cite{Gorbunov:1996:PhysRevLett.76.2495, Gorbunov:POP:1997}: 
\begin{equation} \label{eq:Fr_r}
\vec{B}_\phi(r)\approx
 \frac{mc}{e}\frac{a_0^4\zeta^2 r}{w_0^2} \exp\left(\frac{-4r^2}{w_0^2}\right) \left[1-\frac{4 n_\mathrm{c} c^2}{n(x_2) \omega^2 w_0^2}\left(5-\frac{12r^2}{w_0^2}\right)\right]\,,
\end{equation}
where $\zeta^2 \approx \sin^2(\omega T_l \sqrt{n(x_2)/n_\mathrm{c}})/[4(1-n(x_2)/n_\mathrm{c})^2]$. Owing to the cylindrical symmetry of $\vec{B}_\phi$, the spin of electrons with symmetric trajectories with respect to the wakefield axis precess symmetrically but in opposite direction in the plane orthogonal to $\vec{B}_\phi$, as sketched in Fig.~\ref{fig:scheme}(c). This symmetric precession implies a symmetric spin spreading with $\braket{s_z} \approx \braket{s_y} \approx 0$ [see Fig.~\ref{fig:x250}(c)] together with a decrease of $\braket{s_x}$ [see Fig.~\ref{fig:x250}(d)]. The EBP is therefore obtained by integrating $s_x(r)$ over the electrons of the cylindrical injection volume:
\begin{equation} \label{eq:P}
P = \braket{s_x} = \frac{2}{R_\mathrm{inj}^2} \int_{0}^{R_\mathrm{inj}}{s_x(r)  \, r \D r}.
\end{equation}
In order to carry our the integration in Eq.~(\ref{eq:P}), an estimate of $R_\mathrm{inj}$ is needed. Our simulations indicate that, even though $R_\mathrm{inj}$ weakly depends on the laser intensity, the injection volume radius is roughly $R_\mathrm{inj}\approx w_0/2$ (see below). 

Figure~\ref{fig:scanxf} displays the radial distribution of the injected electrons $\D N/\D r$ and the radius of the injection volume [Fig.~\ref{fig:scanxf}(a)], the $s_x$ distribution and the EBP from PIC simulations (black circles) and from the analytical estimation with Eqs.~\eqref{eq:Fr_r}-\eqref{eq:P} [light blue curve in Fig.~\ref{fig:scanxf}(b)], the electric current $I$ carried by the beam [Fig.~\ref{fig:scanxf}(c)], and the electron energy spectrum $\mathcal{E} \D N / \D\mathcal{E}$ [Fig.~\ref{fig:scanxf}(d)],
where $\mathcal{E}$ is the electron kinetic energy, at $t=400\tau$ as functions of laser amplitude $a_0$. In particular, Fig.~\ref{fig:scanxf}(a) shows that the linear increase in the number of injected particles around the axis is followed by a sharp decrease of $\D N/\D r$ around the injection radius $R_\mathrm{inj} \approx w_0/2$ [see Fig.~\ref{fig:scanxf}(a)]. Note that the number of electrons injected in the beam increases with increasing $a_0$, which is also manifest in the increase of the electric current carried by the beam [see Fig.~\ref{fig:scanxf}(c)]. However, the azimuthal magnetic field increases rapidly with $a_0$ [see Eq.~\eqref{eq:Fr_r}], such that an increase in $a_0$ results also into stronger depolarization [see Fig.~\ref{fig:scanxf}(b)]. 

\begin{figure}[t]
\centering
\includegraphics[width=0.48\textwidth]{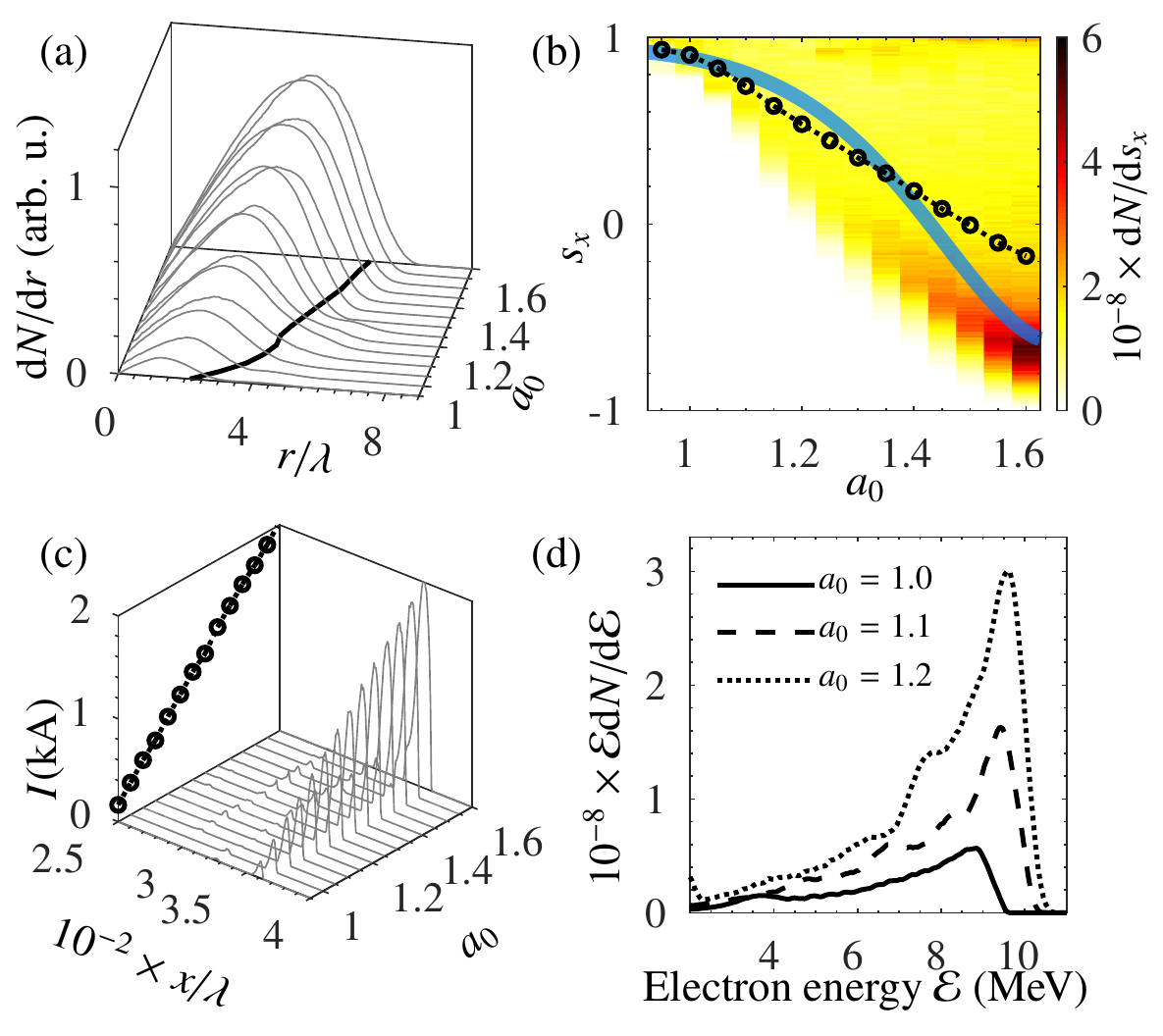}
\caption{(a)~Radial distribution of the electrons of the injection volume as a function of $a_0$. The solid black line corresponds to the radius of injection volume defined as the value of $r$ at which $\D N/\D r$ drops to half its maximum. (b)~EBP at $t=400\tau$ as a function of $a_0$. The light-blue line corresponds to the estimation with Eqs.~\eqref{eq:Fr_r}-\eqref{eq:P} and $R_\mathrm{inj}=w_0/2$. The black circles report the EBP obtained from 3D~PIC simulations. (c)~Electric current longitudinal distribution at $t=400\tau$. The black circles correspond to the peak of the current. (d) Energy spectra of the generated electron beam at $t=400\tau$ driven by optical laser pulses with $a_0 = \text{1, 1.1, and 1.2}$.}  \label{fig:scanxf}
\end{figure}

Figure~\ref{fig:scanxf} shows that in the weakly nonlinear wakefield regime electron beams carrying currents of the order of one kiloampere and retaining the IEP of the plasma can be produced. The final EBP and current from 3D PIC simulations with $a_0 = (1,\,1.1,\,1.2)$ are $(90.6\%,\,73.9\%,\,53.5\%)$ and $(0.31,\,0.59,\,0.90)$~kA, respectively. The difference between the PIC simulation results and the prediction shown in Fig.~\ref{fig:scanxf}(b) for $a_0>1.4$ is ascribed to the limited validity of Eq.~\eqref{eq:Fr_r} for $a_0$ significantly larger than unity~\cite{Gorbunov:POP:1997}.
After the acceleration in the shock-front injector, which occurs within hundreds of micrometers~\cite{PhysRevSTAB.13.091301, PhysRevLett.110.185006, swanson17, Xu:POP:2017}, the accelerator stage can be further extended to obtain GeV beams via multiacceleration-stage techniques powered by the same~\cite{Kim:2013:PhysRevLett.111.165002} or different~\cite{Steinke:Nature:2016} laser pulses. In fact, further acceleration of already ultrarelativistic polarized electron beams weakly alters the EBP~\cite{Vieira:2011:PhysRevSTAB.14.071303}.

\begin{acknowledgments}
We would like to thank M.~B{\"u}scher and A.~Di~Piazza for the helpful discussions. M.~W. thanks Y.~Fu and C.~Lin for their kind help and stimulating discussions about the plasma target and the laser system.
\end{acknowledgments}

\addtolength{\textheight}{10pt}

\bibliography{wakefield}

\clearpage

\end{document}